\documentstyle[aps]{revtex}

\input{epsf}
\begin{document}

\title{
A Search for Disoriented Chiral Condensate 
at the Fermilab Tevatron}
\author{T. C. Brooks,\footnote{Now at Department of Physics, Stanford
University, Stanford, 
CA 94305}\ M. E. Convery,\footnote{Now at The Rockefeller University, New York, 
NY 10021} 
W. L. Davis, K. W. Del Signore,\footnote{Now at Department of Physics,
University of Michigan, Ann Arbor, MI 48109-1120}
T. L. Jenkins, E. Kangas,\footnote{Now at Department of Physics, 
Massachusetts Institute of Technology, Cambridge, MA 02139} 
M. G. Knepley,\footnote{Now at Department of Computer Science, 
University of Minnesota, Minneapolis, MN 55455} K.~L.~Kowalski, and 
C.~C.~Taylor}
\address{Department of Physics, Case Western Reserve University,  Cleveland,
Ohio 44106-7079}
\author{S. H. Oh and W.D. Walker}
\address{Department of Physics, Duke University, Durham, North Carolina
27708-0305}
\author{P. L. Colestock, B. Hanna, M. Martens, and J. Streets\footnote{Now at
Lucent Technologies}}
\address{Fermi National Accelerator Laboratory, P.O. Box 500,  Batavia,
Illinois 60510}
\author{R. C. Ball, H. R. Gustafson, L.~W. Jones, and M. J. Longo}
\address{Department of Physics, University of Michigan, Ann Arbor, Michigan
48109-1120}
\author{J. D. Bjorken}
\address{Stanford Linear Accelerator Center, Stanford, California  
94309}
\author{A. Abashian and N. Morgan}
\address{Department of Physics, 
Virginia Polytechnic Institute, Blacksburg, Virginia 24061-0435}
\author{C. A. Pruneau}
\address{Department of Physics and Astronomy, 
Wayne State University, Detroit, Michigan 48202}

\date{\today}
\maketitle
\begin{abstract}

We present results from MiniMax (Fermilab T-864), a small test/experiment at the
Tevatron designed to search for the production of disoriented chiral
condensate (DCC) in $p - \bar p$ collisions at $\sqrt{s} = 1.8$ TeV in the 
forward direction, $\sim 3.4 < \eta < \sim 4.2$. 
Data, consisting of $1.3 \times 10^6$ events, are analyzed using the robust
observables developed in an earlier paper.  
The results are consistent with generic,
binomial-distribution partition of pions into charged and neutral species.
Limits on DCC production in various models are presented.

\end{abstract}

\section{Introduction}\label{sec1}

The 
purpose of the 
MiniMax test/experiment (T-864) at the Fermilab Tevatron 
as set out in its proposal was to:
(1) demonstrate the feasibility of operating spectrometers in the hostile
environment of the far-forward, small angle
region in high-energy hadron colliders; (2) search for the presence of 
disoriented chiral condensate (DCC) and possibly related exotic phenomena
such as Centauro events; and (3) contribute data on inclusive spectra and
multiplicity distributions in an unexplored region of 
phase space \cite{earlyMiniMax,MThesis}.  
The experiment was proposed in April 1993,
commissioned by January 1994, and upgraded in several stages during the next
two years.  The data reported here were acquired in January 1996.

The principal purpose of this paper is to report the results of our DCC
search. A signal for the
formation and decay of disoriented chiral condensates \cite{earlyDCC,dccrevs} 
in hadronic and
heavy-ion collisions is an anomalous joint multiplicity distribution of neutral
and charged secondary pions, reflected in the probability density
\begin{equation}
P(f)_{DCC} = {1\over{2 \sqrt{f}}},	\label{1byrootf}
\end{equation}
where $f$ is the fraction of the total number of pions which are neutral. 
(There are a variety of proposed mechanisms
other than DCC which might also lead to this distribution
\cite{othermechs}. We will not hereafter
explicitly make this distinction.)
Note that the distribution Eq. (\ref{1byrootf})
differs markedly from the  ``generic", binomial partition of
pions into charged and neutral species expected from
ordinary production mechanisms.   

Neutral pions were not reconstructed in this experiment.  Instead, we
studied the joint multiplicity distribution of charged particles and gamma
rays.  In a recent publication \cite{robust}, we showed that robust
observables can be constructed from such data which still contain
much of the information in Eq. (\ref{1byrootf}) regarding the presence
(or absence) of DCC.  It is this method which we apply to the MiniMax data.

The basic detector requirements of a DCC search, then, 
are to be able
to count, event by event, the number of charged particles and photons in
a given acceptance.  The detailed design of the MiniMax
detector was determined by a variety of considerations.
The far-forward direction of production angles less than
$\sim 50$ mrad was chosen because cosmic ray data provide hints of novel 
phenomena
in this region of phase space, and because it is largely unexplored
at hadron colliders.  In this region it is necessary to determine carefully the
production angles of charged particles and the conversion products of photons.
We therefore designed a forward spectrometer with a large number of planes
of multiwire proportional chambers (MWPC's).  A plane of Pb converter located
within the spectrometer permitted the identification of photons through
their conversion products.  An electromagnetic calorimeter placed behind the
spectrometer provided additional information.

The acceptance of the spectrometer was quite small.  This was due to a 
combination
of fiscal and physical constraints, together with theoretical considerations 
regarding
the possible size of DCC domains.  Available resources dictated the choice of 
MWPC's as the detector technology.  Once this choice had been made, the
cramped environment of the detector dictated the acceptance:  there was no 
space for additional detection elements.  The resulting fiducial region of
the MiniMax detector was only 
$\Delta \eta \Delta \phi \approx 0.75$. 
(It is amusing to note that
this 
corresponds to about $0.75$ steradian
in a reference frame in which the axis of the detector is boosted to $90$ 
degrees with respect to the beam axis.)
Nevertheless,
this choice of acceptance is consistent with the consideration 
that the correlation length of the DCC chiral 
order parameter may be so small that even large acceptance detectors
should be subdivided into cells of the order of the MiniMax acceptance
in order to avoid averaging out the DCC effects.

The detector elements were modular and portable which permitted efficient 
maintenance,
modifications and upgrades during the extremely limited periods of access
to the detector.  In the development period from 1993-1996, the number of 
MWPC's
was increased from an initial 8, first to 12, then to 16, and finally to 24 
planes as
the need for additional redundancy became clear.  The entire detector was 
removed
and rebuilt three times during this period.  Early running demonstrated the 
need for
a special beam pipe to improve resolution and decrease backgrounds.  This beam
pipe was commissioned in 1995.  The original choice of large-angle stereo 
for the
MWPC's was also changed in 1995 to small-angle stereo in order to improve 
pattern recognition capability, thereby reducing the number of spurious track 
candidates though with some loss of resolution.

It was also recognized during the development period that it was possible 
to add
simple detection elements in 
the opposite 
hemisphere which
provided tags of the presence of leading or diffractively produced 
(anti)baryons in
each event.  These tagging detectors were installed and tested in 1995 
and worked
with demonstrably high purity during the data runs.

The detector, as it was configured during the data runs reported here, is 
described in
more detail in the next section.  Section \ref{sec3} describes the
 analysis chain from track finding through
the determination of the joint multiplicity distribution of charged 
particles and
converted photons.  Section \ref{sec4} reviews the use of robust observables
to search for the presence of DCC in the data.   The results are presented in
Section \ref{sec5}.  Section \ref{sec6} contains a summary and our 
conclusions.

\section{The Detector}\label{sec2}

\subsection{Apparatus}

\begin{figure}
\epsffile{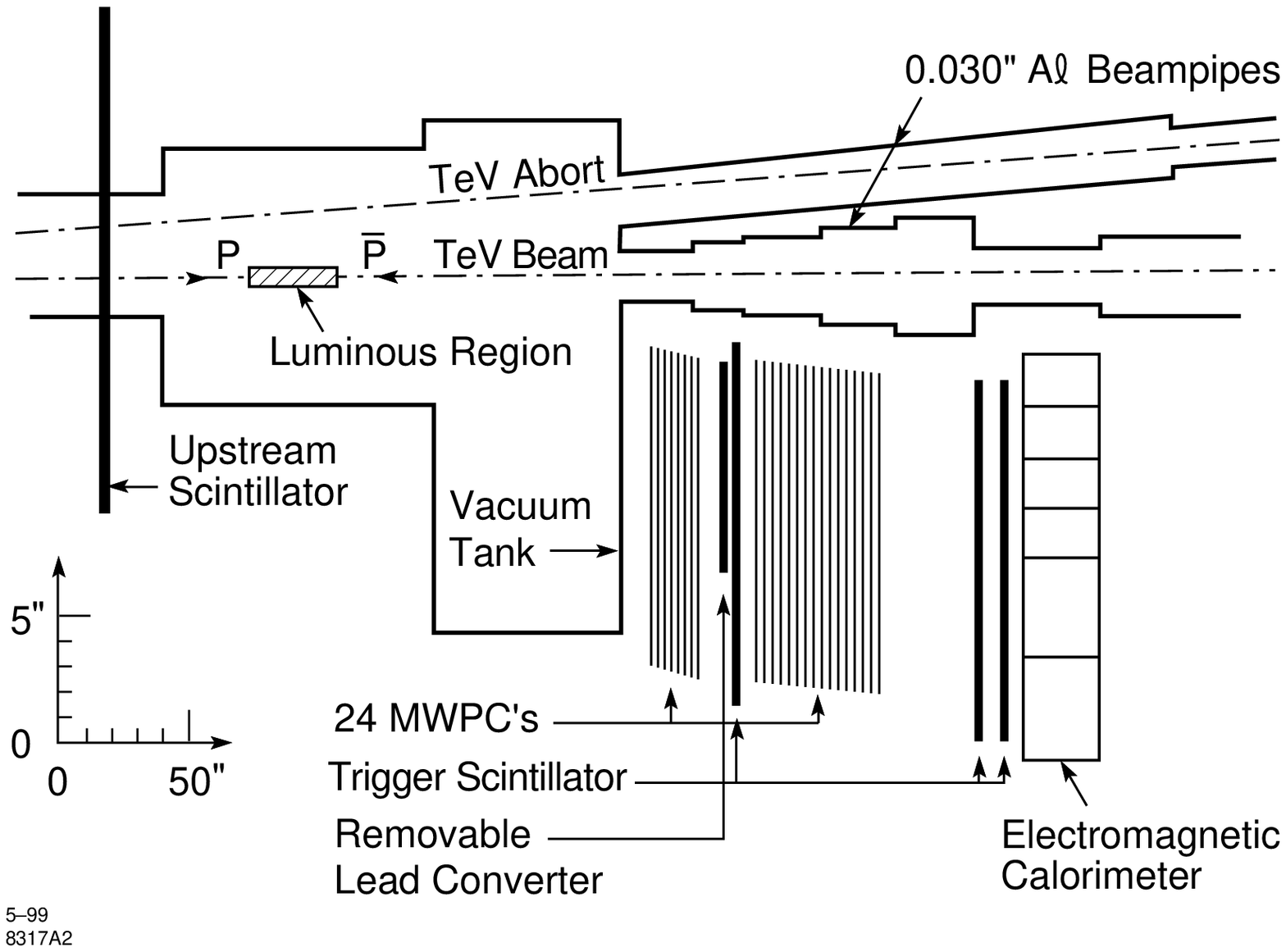}
\caption{Plan view of 
the final configuration of the MiniMax detector, illustrating the 
tracking 
detectors, the beampipe architecture and the location of the trigger 
scintillator
elements.}
\label{Detector}
\end{figure}

\begin{figure}
\epsffile{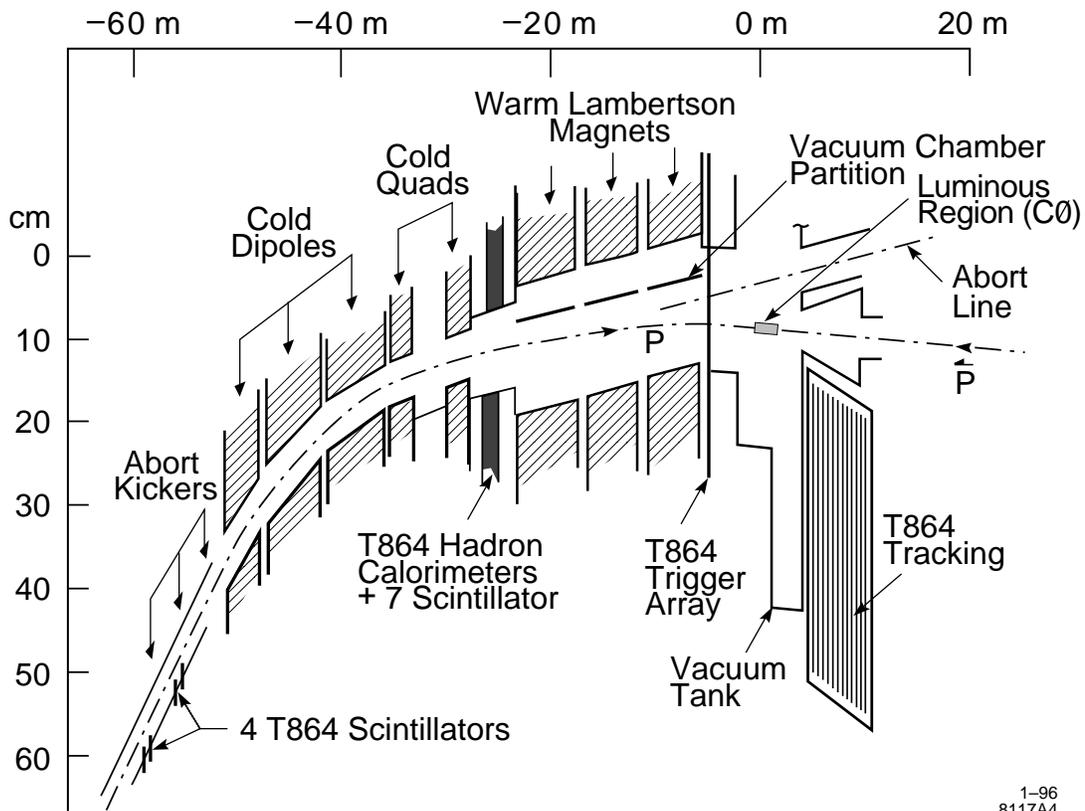}
\caption{Plan view of 
the final configuration of the MiniMax detector illustrating the 
incorporation
of detector elements amidst the Tevatron magnets in order to obtain
leading particle and diffractive tags in the downstream $\bar p$ direction.}
\label{tagdet}
\end{figure}

The MiniMax detector was located at the C0 region of the Fermilab Tevatron.
The final configuration of the detector, used for 
acquisition of the data reported
here, is illustrated in Figures \ref{Detector} and \ref{tagdet}.  The 
salient parts of the
apparatus are the beam pipe, the MWPC tracking telescope including remotely
movable Pb converter, the trigger scintillator, the electromagnetic 
calorimeter, and
the upstream (in the sense of proton motion in the Tevatron) tagging detectors.

Early running during the MiniMax development period confirmed the need for a 
special beampipe.  Two considerations governed the design:  the need for 
a ``thin"
window to minimize interactions of particles before reaching the detector; 
and the
need to minimize backgrounds created by particles outside the acceptance,
particularly at higher $\eta$, interacting with the beampipe and showering.  
The 
design of this special beampipe and vacuum tank was complicated by the fact
that the Tevatron abort system was also located at C0.

The general features of this special beampipe are illustrated in 
Figure \ref{Detector}.
All elements of this beampipe were constructed from aluminum.
The large central vacuum tank insured that there was no material
within 12 cm of the collision point. This tank terminated on the detector 
side in
an aluminum plate, which provided transitions to both the Tevatron beampipe
and the abort pipe.   The MWPC tracking telescope viewed the collision 
region through a circular window  22.86 cm in diameter and 0.64 cm
in thickness milled into this plate.
In the region of the MWPC's, the Tevatron beampipe was flared in a stepped
cylindrical fashion using aluminum tubing of
varying radii and thickness 0.76 mm. 
In this way, interactions of forwardly produced particles with
the beampipe were localized in pseudorapidity.

Tracking information was provided by 24 multi-wire proportional chambers 
(MWPC's).  Each chamber had 128 wires with spacing 2.54 mm and an active
area of approximately 32.5 cm x 32.5 cm.  
Their construction was similar to those described by Bevington et al.
\cite{Bevington}.
The chambers operated on a mixture
of 80\% Ar and 20\% $\rm CO_2$.  Two types of readout electronics were used:
half the chambers were read out with latch-only information, and the other half
recorded the magnitude of the charge deposited on each wire. 

The forward eight chambers served to identify charged particles coming from
the collision vertex.  This was followed by 1 $\rm X_0$ of Pb converter which
could be moved by remote control into and out of the detector acceptance.  This
was followed by 16 more MWPC's which served to identify gamma ray conversion
products and improve the resolution of charged particles.  

The MWPC planes 
were perpendicular to the beam line and oriented so as to provide 
three-dimensional positions of all tracks traversing the detector
in a single event.  Considerable effort went into optimizing the orientation 
of the chambers with  respect to redundancy, resolution and pattern recognition.  In
describing the final orientation of the MWPC's, it is convenient to introduce
a coordinate system in which the $z$-axis is along the Tevatron beam, the 
$u$-axis is perpendicular to the beam and points towards the centerline of the
detector, and the $v$-axis is orthogonal to the other two.  Three of the eight
chambers located in front of the Pb converter were oriented with their wires
parallel to the $v$ axis, two at $\pm 15^\circ$ to this direction, and 
the remaining
three with their wires at $\pm 15^\circ $ and parallel to the $u$ direction.  
Behind
the Pb converter, every other chamber was oriented with its wires parallel
to the $v$ axis.  The remaining chambers were oriented at various angles  but
always within $\pm 15^\circ $ of the $v$ direction.  This arrangement, with
eleven of the twenty-four chambers having parallel wires, permitted simple
visualization and formed the basis for one of the two track reconstruction 
algorithms
used for analysis.

The array of twenty-eight lead-scintillator electromagnetic calorimeter
modules, located behind the MWPC tracking system, provided information
on photons and showering charged particles traversing the apparatus.  This
information has been useful for a variety of diagnostic purposes.  Pattern
recognition proved to be difficult, however, because of the rather coarse
angular resolution and the rather large background levels.  The data from
this system have not been directly incorporated into the analysis presented
here.

Figure \ref{tagdet} illustrates the detectors installed within the 
Tevatron
lattice in order to tag leading particles and diffractive events. 
The presence of
the Lambertson abort magnets ensured that particles leaving 
the collision area in the direction opposite
the main detector would traverse a magnetic field within a comparatively large
aperture beam pipe.  Fortuitously, there was approximately two meters 
of free space between
the Lambertson magnets and the quadrupole magnets in which it was possible
to position two 10 cm x 10 cm x 117 cm
lead-scintillator hadron calorimeter modules, one on each side of the beam pipe.
One module thus saw zero-degree neutral particles from the collision
region, albeit
with a small acceptance and through approximately one interaction length of 
material.
Forward-produced negatively charged particles of $x \sim 0.5$ were bent into
the other module.  Several pieces of scintillator provided additional 
information
useful in characterizing these events.
Four scintillation counters were also placed adjacent to the beam pipe 
in the vicinity of the abort kicker magnets, approximately 60 m 
upstream.  The magnetic architecture of the C0 straight section was such that
antiprotons of $x \sim 0.9$ exited the beampipe in this region.  
The scintillation
counters detected the resulting showers.

\subsection{Operating Conditions, Trigger and Rates}

A hodoscope composed of eight scintillation counters arranged in a square array
47 cm on a side and 2.5 cm thick with a 16.5 cm square aperture was centered on 
the 15.24 cm beam pipe at z= -194 cm.  (The location of this hodoscope is
indicated by ``upstream scintillator" in Figure \ref{Detector}.)
Immediately behind the Pb converter,
two scintillation counters, each 20.3 cm x 40.6 cm x 1.27 cm were mounted 
together
to form a square region covering the acceptance of the MWPC's.  Two additional
arrays of the same size, were mounted immediately behind the final MWPC.
The trigger required a coincidence of hits with appropriate timing in 
the hodoscope, in
the array immediately behind the converter, and in the final array, together 
with a beam-crossing 
time signal provided by the accelerator.  The
experiment was gated off during Main ring acceleration periods because of high
backgrounds.

The trigger cross section was quite large despite the small acceptance of the
detector,  owing to the large amount of material near the collision 
point
(such as the Tevatron beam pipe, the Main Ring, the floor, etc.) which 
efficiently
generated secondary shower products.  The actual trigger cross section was
estimated to be 43 mb.  This was determined from the CDF measured inelastic
cross section \cite{CDF}, the real-time observed D0 luminosity, the MiniMax
trigger rate, and the ratio of the $\beta$ functions between the 
C0 and D0 collision
regions, including corrections for finite bunch length and bunch-to-bunch
intensity variations.  Detailed GEANT simulations of the detector and its 
environment \cite{MThesis}
were consistent with the observed trigger cross section.  
A small fraction
of the events
($\sim 1.4\% $) were associated with high-mass diffractive dissociation of the
antiproton.

Under ordinary operation during run IB of the Tevatron, the beams at C0 were
separated by electrostatic separators, and no collisions occurred in the
C0 collision area.  
MiniMax commissioning was done during special runs, typically at the ends of
stores when the beams were scraped down to sufficiently low intensities to permit
the separators to be turned off.  The data reported here were taken during six
days of special low-luminosity running during January 1996.  
Data, consisting of $1.3 \times 10^6$ events taken with the
Pb converter in the acceptance of the detector, are analyzed in the remainder
of this paper.  
The luminosity at the C0 collision point 
was inferred from the D0 luminosity corrected for
differences in the magnetic architecture at the two points and the fact
that bunches that collide at C0 are not the same pairs that collide at
D0. The C0 luminosity
during these runs ranged from about $10^{26}$ cm$^{-2}$ s$^{-1}$ to about 
$10^{28}$ cm$^{-2}$ s$^{-1}$.
Triggers 
in these runs occurred at rates from a few Hz up to about 75 Hz.  Due to dead
time, events were only recorded at about $40 \%$ of the trigger rate at the
higher luminosities; essentially every event was recorded at the lower 
luminosities.
Additional data, 
including data with the Pb converter outside of the acceptance,
were also taken during this period, but are not directly used in this analysis.

\subsection{Backgrounds}

The detector was located very close to the beam line and as a consequence
events typically showed many reporting wires in the MWPC's.  The median
number of reporting wires per event was 210, or 6.8\% of the total number
of wires.  
The background was not uniformly distributed; wires closer to the beam
had a higher probability of reporting than wires farther away.  The distribution
of reporting wires could be modeled reasonably well by asuming that for any
single event, the density per unit area of ionizing tracks 
was $A/r$, with $r$ 
the perpendicular distance from the beam.  The parameter A varied from 
 $0.12$ $ \rm cm^{-1}$ in the front chamber 
 to $0.31$ $ \rm cm^{-1}$ in the rear
chamber.
Most of these hits were not associated with tracks from the 
collision vertex.  Visual examination of the events showed that in many 
cases, multiple tracks entered the chambers from the adjacent beam pipe.  
Not all reporting wires were obviously associated
with tracks.

Operation under a broad range of conditions provided clear evidence that these
backgrounds were correlated with secondaries produced  in the proton-antiproton
collision of interest, and were not associated with beam gas interactions.
In the runs reported in this paper, the beam-gas trigger rate was always
less than 5\% of the collision trigger rate.

The GEANT simulations of the experiment and its environment
reproduced the general form of the
distribution of hit wires, but consistently underestimated their number giving
values that were about 60\% of the actual number in the front of the
spectrometer to about 50\% in the rear.  Great effort was
expended, without much success, in attempting to understand the discrepancy.
Similar discrepancies have been noted in other uses of similar chambers, 
where it has been suggested that the discrepancy is due to protons knocked out
of the mylar window of the MWPC by very low energy neutrons, which may
not be well
modeled by GEANT \cite{ALICE}.  

\section{Joint Multiplicity Distributions of Charged Particles and
Converted Photons}\label{sec3}

\subsection{Algorithms for Finding Charged Particles and Photons}

The output of the MWPC's is, for each event, a list of wires reporting hits.
>From this it is necessary to reconstruct the  number of charged particles
and converted gamma rays 
within the acceptance of the detector
arising from the proton-antiproton collision.  There
are two principal stages of this analysis:  the identification of track segments,
and the matching of track segments at the Pb converter in order to identify
converted gamma rays and through-going charged tracks.

While both tasks are essential, the first, the identification of track segments,
is particularly critical because of the large number of background hits.  
In particular, at very high occupancies, spurious tracks arising from the
accidental juxtaposition of hit wires became a serious problem.  In practice,
events with more than 600 wires (20\%) reporting have been eliminated from the
analysis reported in this paper for this reason.  
This is approximately 3\% of the total data set.
In order to assess systematic effects arising from these considerations,
MiniMax developed two distinct trackers.  While both algorithms are
rather intricate in their details, it is worth briefly reviewing the basic principles
underlying each of them.

The principal tracker,
which we will refer to as the
``combinatorial tracker" in the remainder of this
paper, used various combinations of four non-parallel 
``cross-hair" chambers
to define candidate tracks.  
The other chambers were then searched for confirming
hits.  If the resulting collection of wire hits satisfied suitable track quality criteria,
it was declared a track segment, with parameters determined by a least-squares
fit.

The second track reconstruction algorithm,
which we shall refer to as the ``$u-v$" tracker in the
remainder of this paper, exploited the power of the 11 chambers
with parallel wires (in the $v$-direction).  Track candidates were first identified
in this projection.  The algorithm then examined the perpendicular plane through
the candidate track, checking for candidate trajectories in this projection, which
had significantly poorer resolution due to the small angle stereo in the rear chambers.
If suitable track quality criteria were satisfied, 
the parameters of the track that passed through the sensitive area of
those wires identified with the track
were determined by a simplex linear programming
algorithm.
We will present selected results from this track reconstruction
algorithm to illustrate
systematics; our principal results are based on the combinatorial tracker
unless otherwise noted.

The output of the track reconstruction algorithms were then used to determine the 
the number of charged particles and converted gamma rays from the interaction
region entering the fiducial region of the MWPC tracking telescope.  Charged
tracks were identified as track
segments in the front chambers that matched track segments in the rear 
chambers at the Pb converter.  One or more tracks in the rear
chambers appearing to emerge
from a common point on the Pb converter, without a matching track segment
in the front chambers, were identified as conversion products, and the
group was counted as a converted gamma ray.  
The algorithm includes a
number of cuts which were developed and which are described in detail in
\cite{MThesis}. 

One point worth noting is that the algorithms for
vertex fitting at the converter were developed after the bulk of the
work on the combinatorial tracker had been completed, and the various
cuts were developed on the basis of the output of that tracker, on both
real data and the Monte-Carlo to be described shortly.  The $u-v$
tracker was developed after the vertex algorithms were mature, and no
effort was made to retune the various cuts in these algorithms.

\subsection{Algorithm Performance}

The assessment of the algorithms for track reconstruction and vertex
fitting at the converter is  complicated by the fact
that the Monte-Carlo significantly under-estimated the density of hit
wires, as described above. This means that estimates 
based on Monte-Carlo simulations of efficiencies
for identifying the various species of particles 
cannot necessarily be quantitatively trusted.  Fortunately, as described
in the next section, the actual MiniMax DCC search relies on techniques
which are insensitive to these quantities.  Nevertheless, it is worthwhile
to briefly describe the performance of the algorithms as understood from
the Monte-Carlo.

The details of our use of the standard PYTHIA event generator
\cite{PYTHIA} and of the GEANT detector simulation package
\cite{GEANT}, together with a description of our Monte-Carlo
DCC generator have been
described elsewhere \cite{robust,MThesis}. Here we discuss using these
tools to assess how well our algorithms identify tracks that represent
charged particles or converted photons originating in the primary
collision, as distinguished from secondary and spurious, i.e., fake,
tracks.

The efficiencies are calculated by comparing reconstructed charged
tracks and photons with those known to be present in the PYTHIA-GEANT
events.  The actual charged tracks are defined as charged particles
from PYTHIA which are aimed into the acceptance, while converted
photons are defined in terms of $e^{\pm}$ tracks in GEANT which
originate in the region between the front and rear groups of MWPC's,
hit at least 14 of the 16 rear MWPC's, and can be matched to a 
PYTHIA photon.  A charged track or photon is said to be found if
one is reported in the tracking output within a small distance of
the actual track at the converter plane. 

Reconstructed tracks that do not match up with actual tracks are
declared to be fakes.  A minimum bias sample of PYTHIA-GEANT events
was scanned to determine the sources of such fakes.  Most fake charged
tracks were secondary particle tracks from decays.  Particles resulting
from interactions in the detector and its environment were also
significant.  Less than 1\% of the ``fake" charged tracks appeared to be
genuinely spurious.  ``Fake" photons arise for similarly diverse
reasons.  

We now summarize the results of these studies.

\begin{figure}
\epsffile{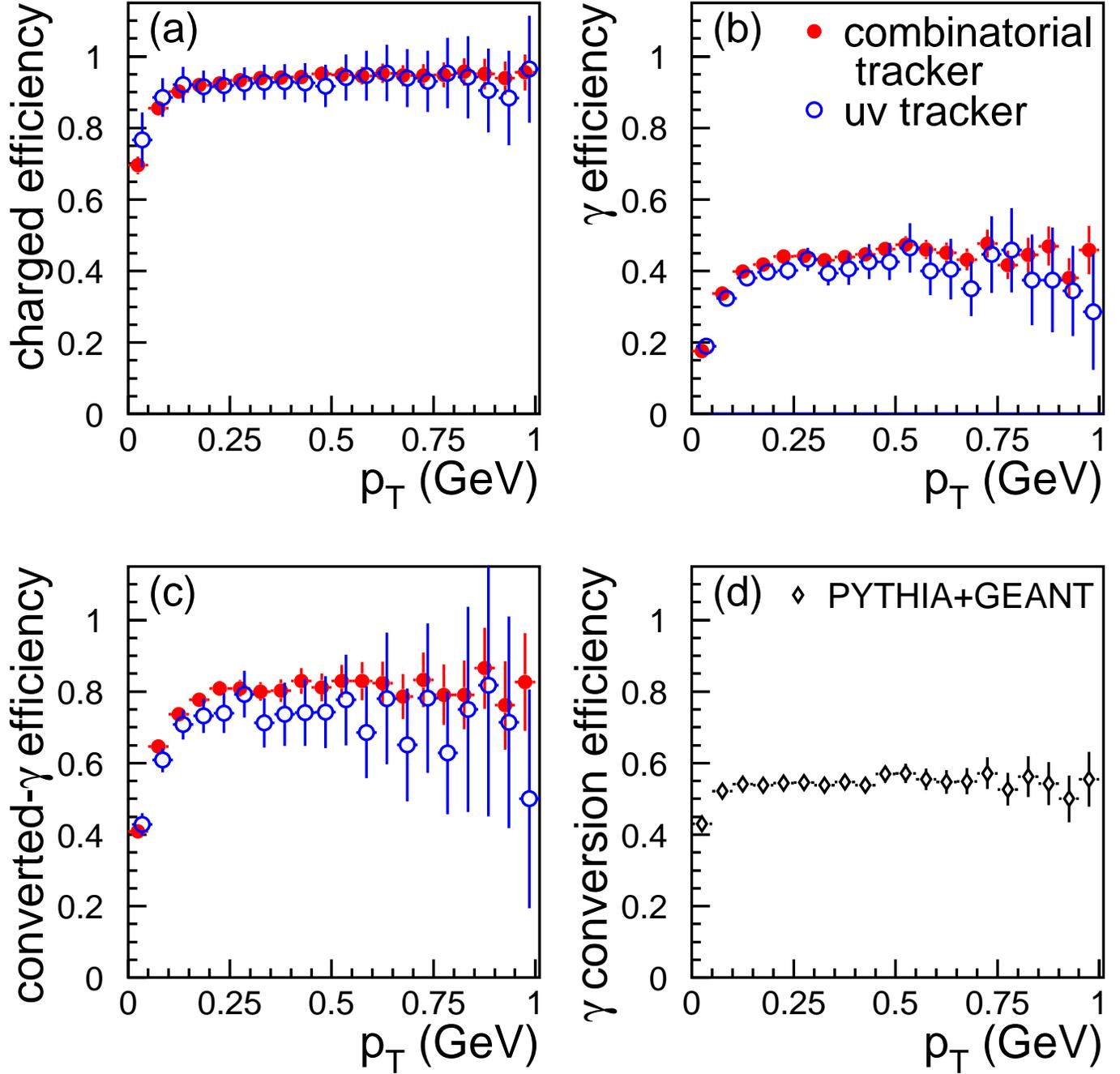}
\caption{PYTHIA + GEANT efficiency estimates for both combinatorial
and $u-v$ track-finding algorithms.
(a) The efficiency for finding charged particles as a function
of $p_t$.
(b) The over-all efficiency for finding photons as a function
of $p_t$.
(c) The efficiency for finding photons known to have converted,
as a function of $p_t$.
(d)  The efficiency for photons to convert, as a function of $p_t$.}
\label{effpt}
\end{figure}

DCC production is often thought to be characterized by low $<p_t>$, so it
is important that we maintain good efficiencies for finding charged
particles and converted gammas in this region. These
efficiencies are plotted in Figure  \ref{effpt}.  Good efficiencies
for both charged particles and photons are maintained to $p_t$
below 100 MeV/c.  

\begin{figure}
\epsffile{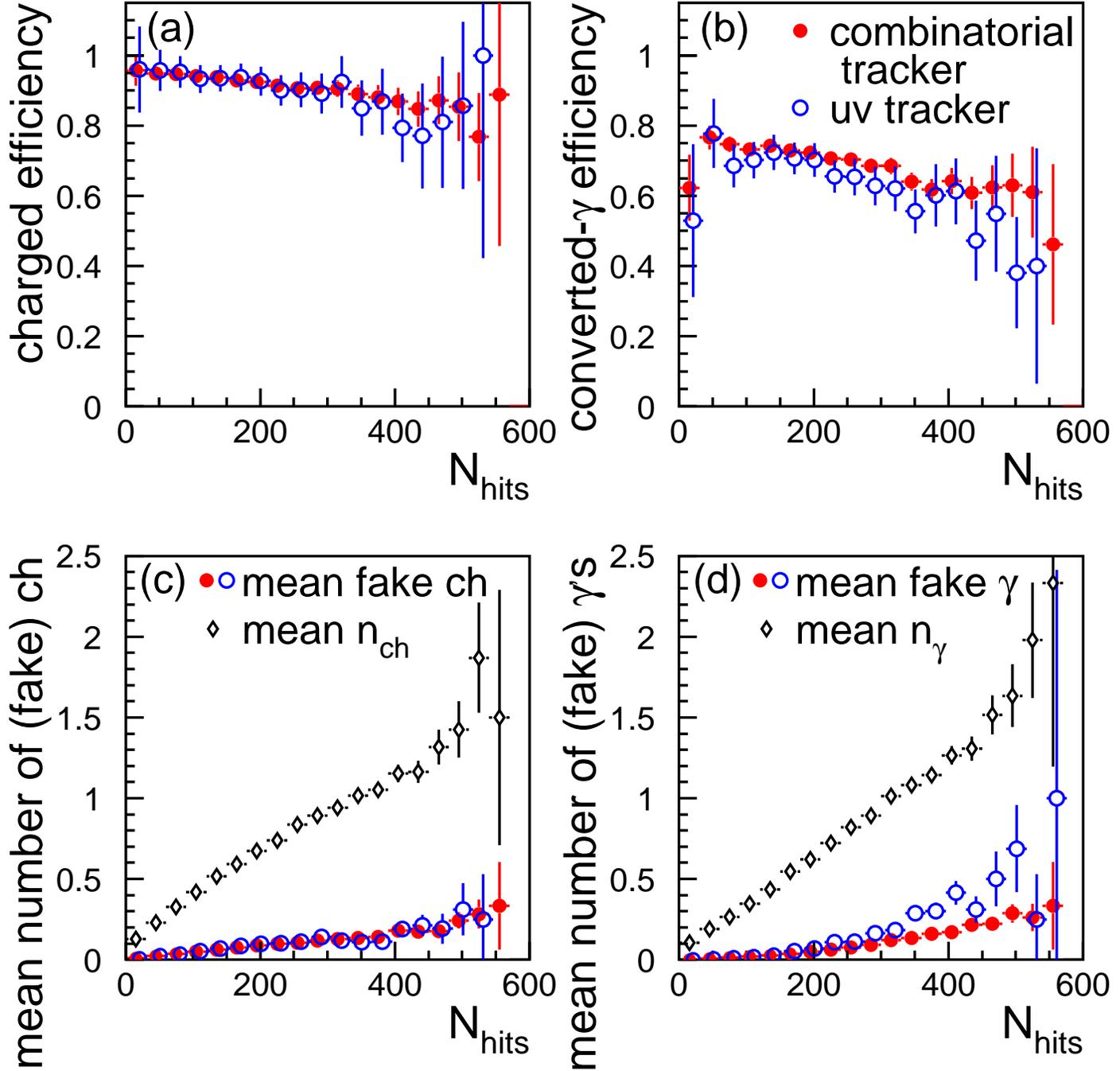}
\caption{PYTHIA + GEANT efficiency 
and ``fake" estimates for both combinatorial
and $u-v$ track-finding algorithms as a function of $N_{hits}$,
the number
of wire hits reported in the event.
(a) The efficiency for finding charged particles as a function
of $N_{hits}$.
(b) The efficiency for finding converted photons as a function
of $N_{hits}$.
(c) The mean number of correctly identified and
``fake" charged particles found, as a function
of $N_{hits}$.
(d)  The mean number of correctly identified and
``fake" converted photons found, as a function
of $N_{hits}$.}
\label{effnhits}
\end{figure}

As noted above, a great concern is the performance of the reconstruction
algorithms as chamber occupancies become large. 
Figure~\ref{effnhits}
illustrates the performance of the algorithms for
correctly identifying charged particles and photons
as a function of $N_{hits}$, the number of wires reporting hits in
the event.  Efficiencies
are high, and the mean number of ``fakes" is significantly smaller
than the mean number of correctly identified particles.  It should
be noted that the $u-v$ tracker is comparable to the combinatorial
tracker in identifying charged particles, but seems to have lower
efficiencies for correctly finding converted photons, and a significantly
enhanced probability of finding ``fake" photons at high occupancies.

\begin{figure}
\epsffile{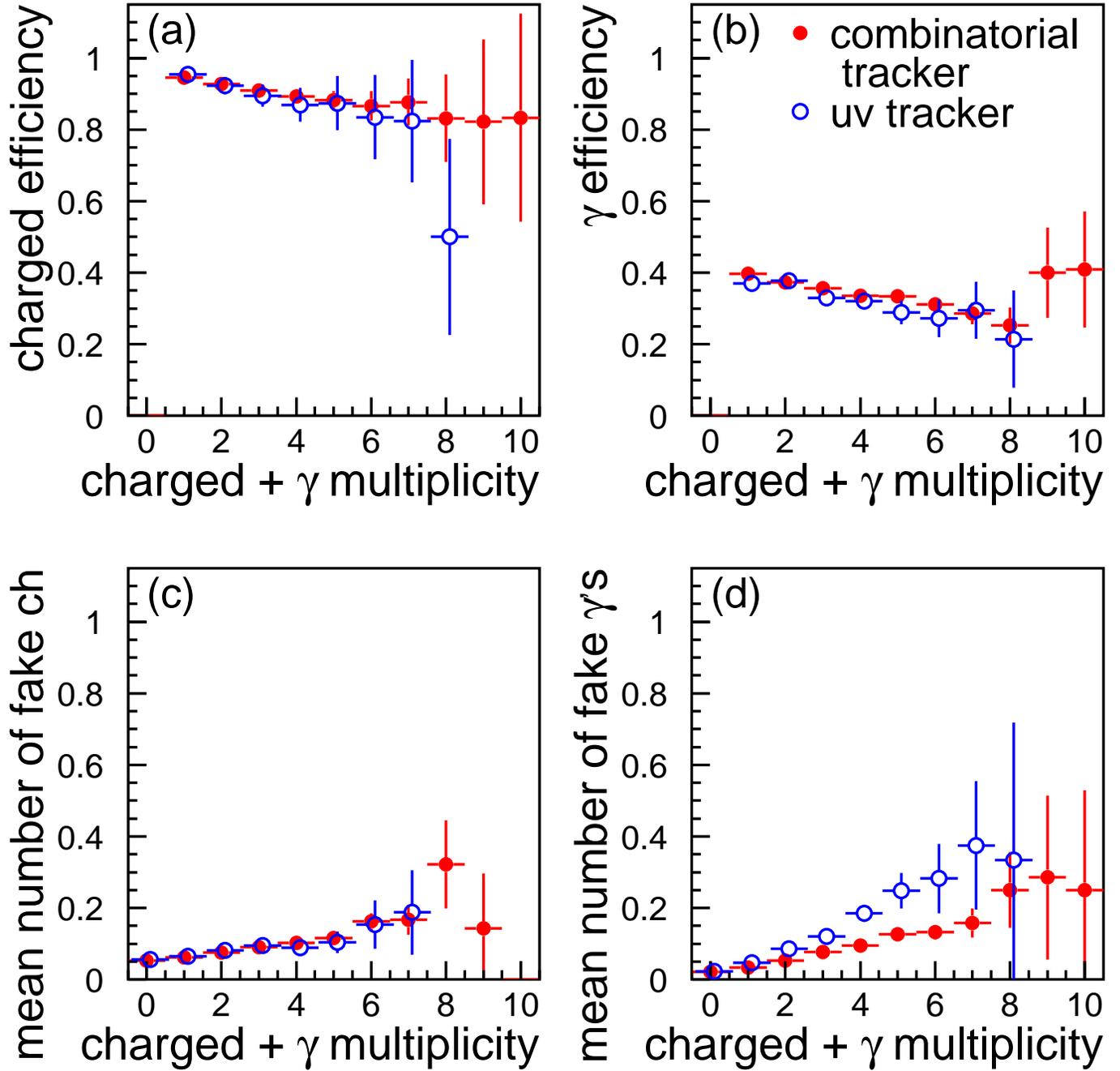}
\caption{PYTHIA + GEANT efficiency 
and ``fake" estimates for both combinatorial
and $u-v$ track-finding algorithms as a function of total multiplicity
into the acceptance.
(a) The efficiency for finding charged particles as a function of total
multiplicity into the acceptance.
(b) The efficiency for finding photons as a function of total
multiplicity into the acceptance.
(c) The mean number of 
``fake" charged particles found as a function of total
multiplicity into the acceptance.
(d) The mean number of 
``fake" photons found as a function of total
multiplicity into the acceptance.}
\label{effmult_a}
\end{figure}

Figure \ref{effmult_a} similarly illustrates the performance of
the algorithms as a function of the total multiplicity into the acceptance.

\begin{figure}
\epsffile{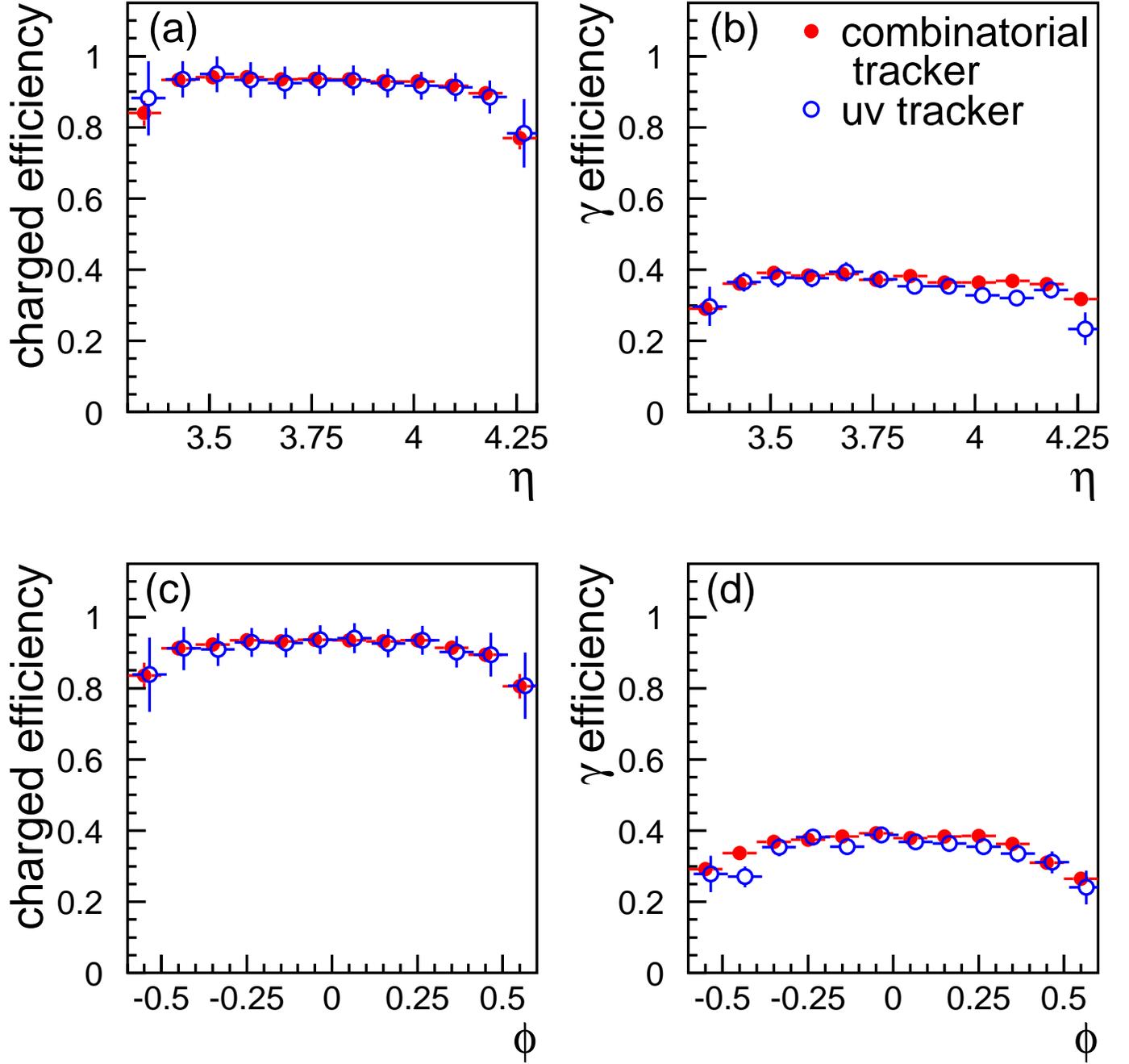}
\caption{PYTHIA + GEANT efficiency 
estimates for both combinatorial
and $u-v$ track-finding algorithms as a function of 
pseudorapidity $\eta$, and azimuthal angle $\phi$.
(a) The efficiency for finding charged particles as a function
of $\eta$.
(b) The efficiency for finding converted photons as a function
of $\eta$.
(c) The efficiency for finding charged particles as a function
of $\phi$.
(d) The efficiency for finding photons as a function
of $\phi$.}
\label{effetaphi}
\end{figure}

As noted above, occupancies in the MWPC's 
vary as $A/r$ and
so are much higher near the beam pipe, that is at higher pseudorapidity,
$\eta$.  Figure \ref{effetaphi} illustrates that the performance of
the reconstruction algorithms is reasonably uniform over the entire
acceptance.

\subsection{Joint Distributions of Charged Particles and Photons}

We now present the basic experimental data of MiniMax:  the observed
joint distributions of charged particles and converted photons within
our acceptance. The data presented is based on the output of the
combinatorial tracking algorithm unless otherwise noted.  We first
present minimum bias data, and then data sets defined by various
tags.

Table \ref{t:ncg_comp} presents the observed minimum
bias joint distribution of
charged particles and converted photons.  This will be the basis
for our DCC search in minimum bias events.

It has been suggested that DCC-like phenomena might occur 
preferentially in diffractive
events \cite{Dino}.  
To test for this, MiniMax identified the subset of the data
in which the scintillation counters in the
vicinity of the kicker magnets at $\sim 60 \; m$ upstream fired,
indicating a leading antiproton with $x\sim 0.9$.  The joint 
multiplicity distribution in this class of events 
(``diffractive-$\bar p$" events)
is presented in
Table \ref{t:ncg_ktag}.

The hadron calorimeter modules located $\sim 25 \; m $ upstream 
permitted MiniMax to isolate a subset  of the data characterized
by leading antiprotons of $ x \sim 0.5$ (forward-$\bar p$ events), 
and by leading zero-degree
neutrals (forward-$\bar n$ events).  
The joint multiplicity distributions are presented in
Table \ref{t:ncg_pbar} and Table \ref{t:ncg_nbar}, respectively.

\begin{figure}
\epsffile{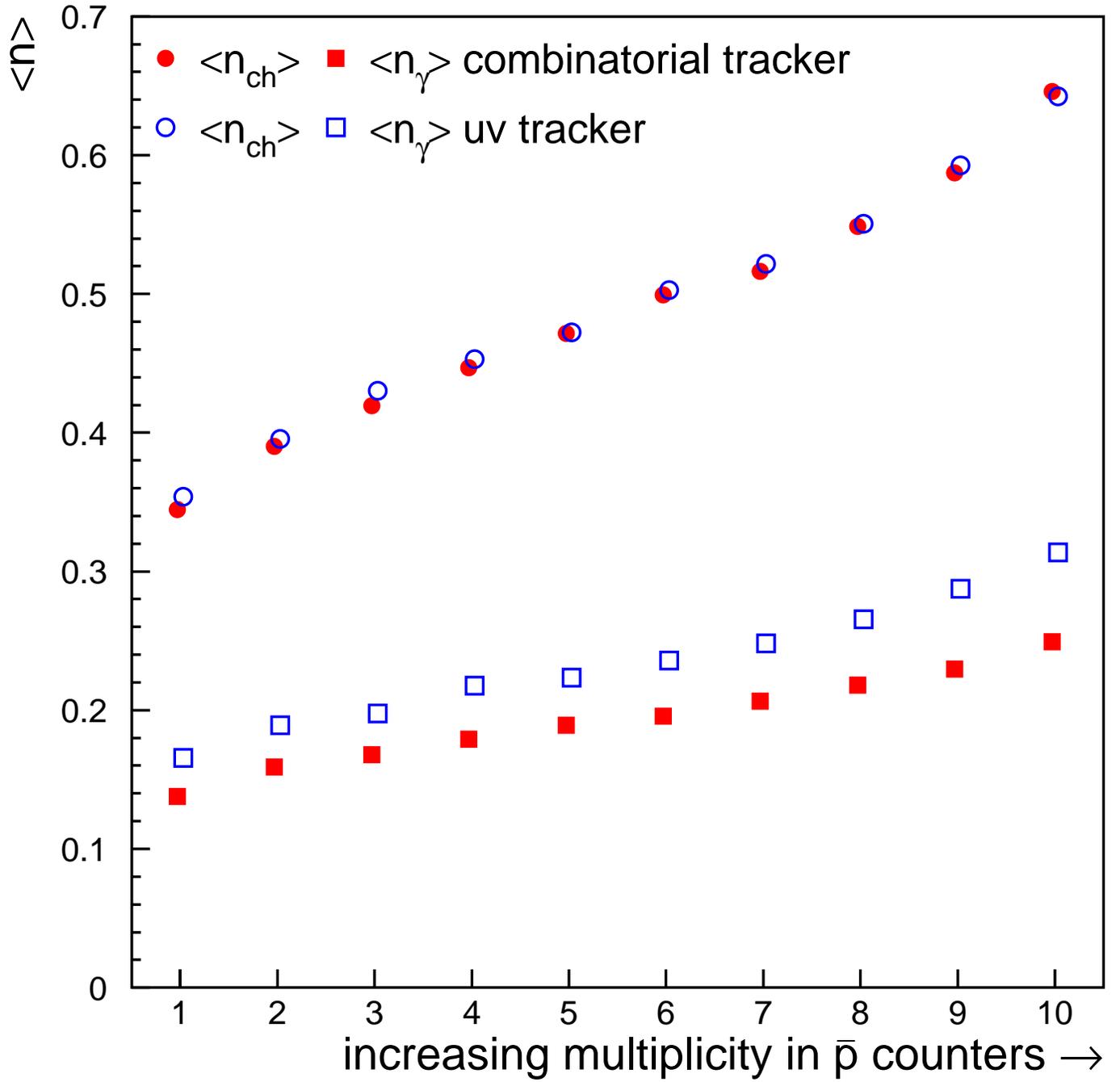}
\caption{Correlations between  multiplicity as seen
by MiniMax tracking and multiplicity seen in the scintillator
hodoscope.  Events were grouped in bins by multiplicity observed
in the scintillator hodoscope such that each bin contains 10\%
of the data.}
\label{pbarcor}
\end{figure}

Many models for DCC formulation suggest that there should be
a correlation with total event multiplicity.  In order to avoid
biasing the data, it is important to cut on multiplicity in a 
manner that does not depend on what is seen in the event within
the MiniMax MWPC acceptance.  Fortunately, the scintillator hodoscope
on the opposite side of the collision vertex provides such a tool.
Peaks in the ADC spectrum of each scintillation
counter provided a calibration
in terms of the number of minimum ionizing particles passing through
the hodoscope.  The data set was then subdivided into ten bins
of increasing multiplicity in the hodoscope, with an equal number of
events in each bin.  Despite being some 7 units of pseudorapidity away from
the fiducial region for tracking, there is a clear correlation between
multiplicity in the hodoscope and the number of particles observed
by the tracking system.  This is illustrated in Figure \ref{pbarcor}. 
Tables \ref{t:ncg_pb1} and \ref{t:ncg_pb10} present the joint 
multiplicity distributions for events with energy deposition
in the hodoscope of less than 2.5 mips 
(minimum ionizing particles) and more than
34 mips, respectively.

Finally, Figure \ref{dndeta} presents the inclusive distributions
$dN_{ch}/d \eta$ and $ d N_{\gamma}/d \eta$, uncorrected for 
detection and trigger efficiencies.  For comparison purposes, 
distributions from PYTHIA, and the output of our entire 
PYTHIA-GEANT-tracking-reconstruction chain are also plotted.  

\section{The DCC Search}\label{sec4}

\subsection{General Strategy}

The output of the analysis chain described in the previous section is, 
for each experimental run, a table of observed numbers of events, 
$N(n_{ch},n_{\gamma})$,
in which $n_{ch}$ charged particles and $n_{\gamma}$ converted gammas 
coming from the collision vertex were
observed within the acceptance of the detector.  The goal of a DCC search
is to use these measured probability distributions to identify, or place
limits on, a component of multiparticle production arising from disoriented
chiral condensates, or other mechanisms leading to anomalous 
charged-particle/photon joint multiplicity distributions.

As should be clear from the preceeding sections, MiniMax faces a number of
challenges in carrying out this analysis.  These include:
\begin{itemize}
\item[(a)] The MiniMax acceptance is small, so that it is improbable that 
both $\gamma$'s from a $\pi^{0}$ enter the detector acceptance.
\item[(b)] The conversion efficiency per $\gamma$ is about 50\%. 
\item[(c)] Not all $\gamma$'s come from $\pi^{0}$'s.
\item[(d)] Not all charged tracks come from $\pi^{\pm}$'s. 
\item[(e)] Because of the small acceptance, the multiplicities are rather low, 
so that statistical fluctuations are very important. 
\item[(f)] Detection efficiencies for charged tracks and $\gamma$'s are 
not the same and are not fully known.
\item[(g)] The efficiency for triggering when no charged track 
or converted $\gamma$ is produced within our acceptance is relatively low and
different
from that for events in which at least one charged particle or converted
$\gamma$ is detected. 
\end{itemize}

Nevertheless, we have identified observables which are robust in the sense
that,
even in the presence of large (uncorrelated) efficiency corrections and
of convolutions from distributions of
produced $\pi^{0}$'s to those of
observed $\gamma$'s, the observables take very
different values
for pure DCC and for generic particle production \cite{robust}. 
Each such observable is a
ratio,
collectively referred to as $R$, of certain bivariate normalized factorial
moments,
that has many desirable properties, including the following:

\begin{enumerate}
\item The $R$'s are not sensitive to the form of the parent pion multiplicity 
distribution.
\item The $R$'s are independent of the detection efficiency for finding
charged 
tracks, provided this efficiency is not,
for example, momentum dependent or correlated with 
other 
variables such as total multiplicity or background level.
\item Some of the $R$'s are also independent of the $\gamma$ efficiencies 
in the same sense as above. In the remaining cases, the $R$'s depend only upon
one parameter,
 $\xi$, which reflects the relative probability of both photons from a
$\pi^{0}$ being 
detected in
 the same event.
\item In 
all cases $R$ is independent of the magnitude of the null trigger efficiency;
see comment (g) above. 
\item The ratios $R$ possess definite and very different values for pure
generic and 
pure DCC pion production.
\end{enumerate}

The idealizations implicit in the realization of properties 1-5 include the
assumptions that
particles other than pions can be ignored, that there is no misidentification
of charged 
particles with photons, and that the production process can be modeled as a
two-step process,
with a parent-pion multiplicity distribution posited, followed by a particular 
charged/neutral partitioning of that population by, e.g., a binomial or DCC
distribution 
function. In addition, there is the vital assumption that detection
efficiencies for finding
a $\pi^{\pm}$ or $\gamma$ do not depend upon the nature of the rest of the
event. 

The validity
of these idealizations, and the utility of the robust observables,
has been studied in the context of the 
MiniMax Monte-Carlo, and their utility is confirmed in this
context \cite{robust}.  
We thus use the robust observables as the basis for our analysis in the
remainder of this paper.  It is important to note, however, that some 
information is lost in this procedure.  While we will be sensitive to the
presence of DCC, we will make no attempt to unfold the parent distribution
of charged and neutral pions, since this would require detailed knowledge
of the detection efficiencies for charged tracks and $\gamma$'s.

\subsection{Factorial Moments and Robust Observables}
In analysis of multiparticle distributions, it is frequently useful to work in
terms of factorial moments, rather than probability distributions.  For
example, in standard analyses of charged particle distributions, rather than
work with $P(N)$, the probability of observing $N$ charged particles, it is
frequently more useful to work with the normalized factorial moments,
\begin{equation}
F_{i} \equiv \frac{\langle N(N-1)\ldots (N-i+1)\rangle} {\langle N\rangle
^{i}}.
\end{equation}
Part of the utility of these variables arises from the fact that they are unity
if the parent distribution $P(N)$ is Poisson, thus essentially removing
statistical fluctuations. 

In order to search for DCC,  the usual multi-particle formalism needs to be
extended to  bivariate distributions.  For the purposes of the MiniMax
analysis, this is given by
\begin{equation}
 F_{i,j}=\frac{\left< n_{ch}(n_{ch}-1)\ldots
(n_{ch}-i+1)~n_{\gamma}(n_{\gamma}-1) \ldots (n_{\gamma}-j+1)\right>}
{\left<n_{ch}\right>^{i}\left<n_{\gamma}\right>^{j}}\ .
\end{equation}

While the normalized bivariate factorial moments are interesting in their own
right, particular ratios, $R$, of them are remarkably robust.  In particular, the
quantities
\begin{equation}
r_{i,1}=\frac{F_{i,1}}{F_{i+1,0}} 
\end{equation}
are robust in the sense outlined above, including independence of both charged
and gamma detection efficiences.
Moreover, it can be shown that for all $i \geq 1$
\begin{eqnarray} 
r_{i,1}(generic) &=& 1, \nonumber \\
r_{i,1}(DCC) &=& \frac{1}{i+1}  ,
\end{eqnarray}
where $generic$ indicates the presence of a binomial distribution, and
$DCC$ indicates a pure-DCC joint probability distribution.  Evidently, the
observables go a long way towards rendering many systematic errors quite
harmless to the DCC analysis.

As noted above, these predictions have been tested in the context
of the MiniMax Monte-Carlo.  The combinatorial tracker yields values
of $r_{1,1}= 1.02\pm  0.02$ when minimum bias PYTHIA events are put
through the full detector simulation and analysis chain.  The combinatorial
tracker also results in a value of $r_{1,1}= 0.58 \pm 0.01$ when run
on a ``pure" DCC, indicating that the various idealizations made in
deriving the robust observables are indeed reasonably robust.

The $u-v$ tracker, developed independently of the vertex-finding
algorithms, has been seen above to have systematic differences from
the combinatorial tracker in its performance.  This is reflected in
the values it yields in the computation of the robust observables, too.
In particular, the $u-v$ tracker yields values
of $r_{1,1}= 1.13  \pm 0.03$ when minimum bias PYTHIA events are put
through the full detector simulation and analysis chain.  The $u-v$
tracker also results in a value of $r_{1,1}= 0.66 \pm  0.02$ when run
on a ``pure" DCC.  Both values are higher than expected, suggesting
that the $u-v$ tracker creates correlations between charged particles
and photons in ways that are not well understood.  It is for this
reason that we report the results of this algorithm, for it provides
an estimate of unknown systematic effects.  Nevertheless, it is
important in what follows to note that the $u-v$ tracker still sees
a significant difference in the values of the robust observables when
DCC is present.

\section{Results}\label{sec5}

\subsection{Minimum Bias Data}

The observed joint frequency distribution for charged particles and
converted photons for the 
1.3 million events tabulated in Table \ref{t:ncg_comp} can be used
to calculate the factorial moments and the $r_{i,j}$.  The results
are tabulated in Table \ref{t:rreal}.  
The lower-order $r_{i,1}$ are close to what is expected
for a binomial distribution  ($r_{i,1}=1$).  
The higher-order ratios  are weighted towards bins of ${\cal
N}(n_{ch},n_\gamma)$  which are statistically limited, and therefore the
deviations from unity are not very significant.   
While the $r_{i,1}$ are robust in the sense defined in section IV, this is
not true of the $r_{i,j}, j>1$ (see \cite{robust}).  
We have nevertheless tabulated these
results for completeness, though they are not useful for the present analysis.
The $u-v$ tracker yields significantly higher values, consistent
with its performance on the Monte-Carlo.
In any case, the ratios are
not smaller than one as would be expected for a contribution from DCC.  
We therefore conclude
that the analyzed events appear to be consistent with production by only
generic mechanisms.

\subsection{Events with diffractive and forward antinucleon tags}

The scintillation counters in the vicinity of the kicker magnets
at $z \sim -60$ m were
used to tag events (``diffractive-$\bar p$" events) 
in which an antiproton of $x\sim 0.9$
was produced and showered into the kicker magnets.
The means 
$\langle n_{ch} \rangle$ and $\langle n_{\gamma} \rangle$ are 
lower for such
events, as would be expected for a diffractive process, where a large fraction of
the total energy is carried away by the beam remnant,  and the charged-charged
and charged-gamma correlations are correspondingly lower.  Table \ref{t:rreal}
reports these values, as well as  the $r_{i,j}$ for the diffractive-$\bar p$ events.   

The upstream hadronic
calorimeters at $z\approx -25$ m are used to tag events with 
$\bar p$'s
with $x_F\sim 0.5$ (forward-$\bar p$ tags)
and small-angle $\bar n$'s. Differences related to
isospin exchange in  diffractive events might be apparent in comparisons
between
events with an $\bar n$ and  those with a $\bar p$. 
The  moments and $r_{i,j}$ are also listed in  Table \ref{t:rreal}.  The mean
number of particles found is higher than that in events with the 
diffractive-$\bar p$,  
but
lower than in the total sample of events, and is lower for the tag on forward 
$\bar n$'s than for forward $\bar p$'s of half the beam momentum, as 
is consistent with energy conservation.   

The $r_{i,j}$ values for these 
diffractive and forward antinucleon tagged events 
do not differ 
significantly from  the values for the total sample.  Therefore, we conclude that
there is no evidence for more DCC  production in events with diffractive 
or forward antinucleon tags, which offers
no support to the conjecture that Centauros are  related to DCC and
are diffractive in nature.   

\subsection{Events with an opposite side multiplicity tag}

As can be seen from
Figure \ref{pbarcor},  there is a long range correlation in
multiplicity which permits us to tag mean multiplicities within the acceptance
of the tracking telescope by cutting on the multiplicity observed in the
scintillator hodoscope even though it is 7 units of
pseudorapidity away.  This provides a powerful tool for checking the hypothesis
that DCC content is multiplicity dependent.  

\begin{figure}
\epsffile{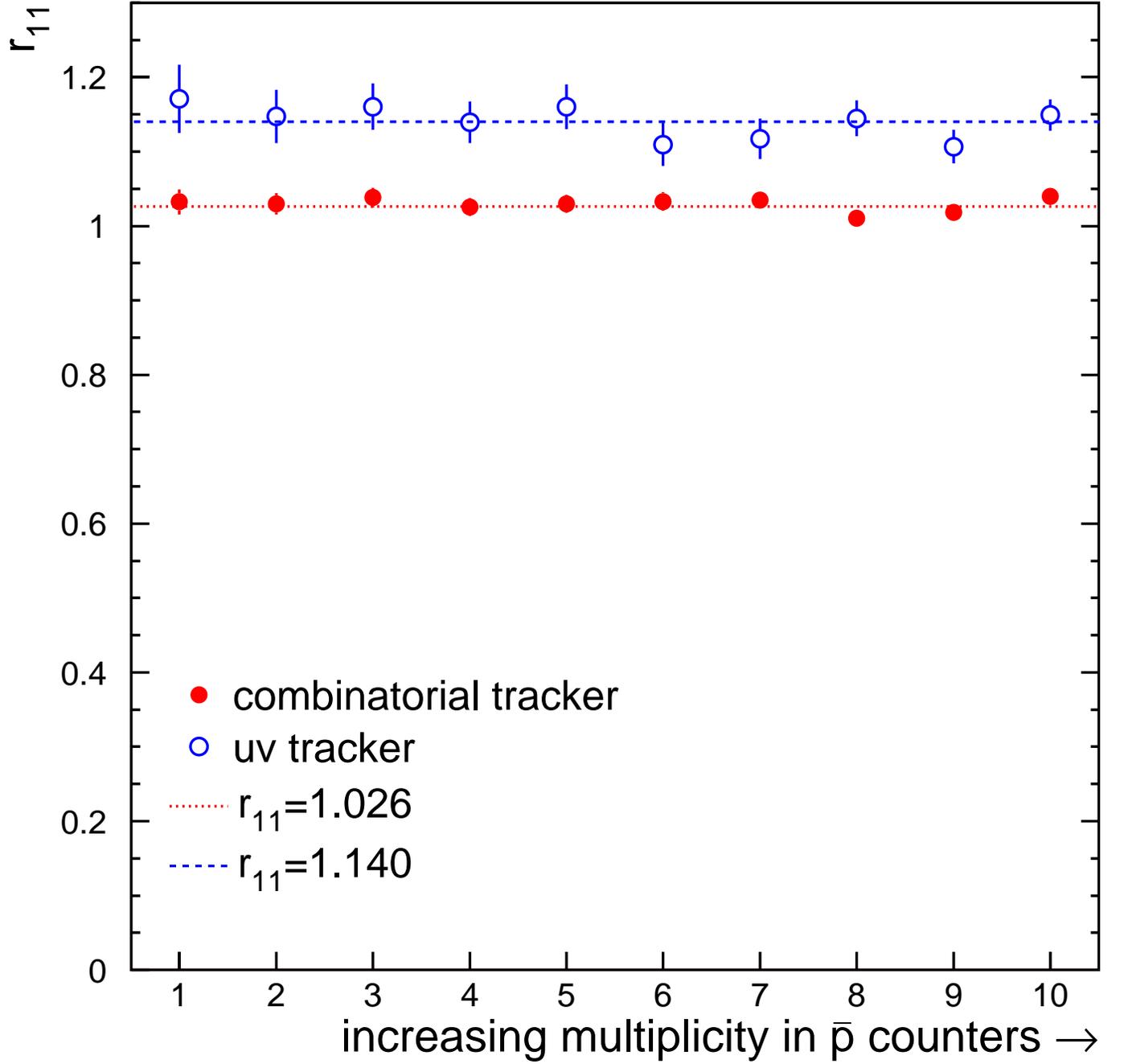}
\caption{The robust observable $r_{1,1}$, for both the combinatorial
and the $u-v$ trackers, as a function of multiplicity in the
scintillator hodoscope.  The multiplicity bins are defined as in
Figure \ref{pbarcor}.  The horizontal lines 
indicate the values of $r_{1,1}$ for 
all events (not separated into multiplicity bins): $r_{1,1}=1.026 \pm 0.004$
for the combinatorial tracking algorithm, and 
$r_{1,1}= 1.140 \pm 0.009$ for 
the $u-v$ tracking algorithm. }
\label{rpbar}
\end{figure}

Figure \ref{rpbar} illustrates the dependence of $r_{1,1}$ on multiplicity.
The horizontal lines in the figure indicate the values of
$r_{1,1}$ calculated from the entire data set.  Testing the hypothesis that
the values of $r_{1,1}$ in each of the ten 
multiplicity bins all come from the same parent value of 
$r_{1,1}$ (that for the entire data set), yields $\chi^2 =7.06$ for the
combinatorial tracker and $\chi^2 =5.58$ for the $u-v$ tracker.  Fitting
to a line gives a small, statistically irrelevant, slope:
$r_{1,1}=(1.0320 \pm 0.0093) - (0.0006 \pm 0.0013)*(\hbox{multiplicity bin \#}) $
with $\chi^2=7.04$ for the combinatorial tracker and 
$r_{1,1}=(1.1563 \pm 0.0227) - (0.0029 \pm 0.0032)*(\hbox{multiplicty bin \#}) $
with $\chi^2=4.62$.  Thus, 
while there is a significant difference between the two
trackers in the absolute value of $r_{1,1}$, neither tracker shows any
indication of a multiplicity dependence in $r_{1,1}$. 

We conclude that there is no evidence
of a multiplicity dependence in the partitioning of pions into charged and
neutral species, and hence no evidence for a multiplicity dependence in
DCC production.

\section{Conclusions}\label{sec6}

The principal goals of test/experiment T864 were (1) to determine whether
spectrometers such as ours could be triggered and would survive the severe
background conditions present in the far-forward direction of a high-energy
collider environment, (2) to search for DCC, (3) to search for exotic phenomena
such as Centauro events, and (4) to provide data on inclusive spectra,
correlations, and multiplicity distributions in a previously unexplored region
of
phase space. 

At the time of the proposal, the first goal of the initiative was
a very serious issue.
There was very little working experience, and what did exist was
not encouraging. But MiniMax ran successfully with many
different detector configurations for a period of about two years.  Much was
learned which should prove useful for the operation and design of future
detectors
in the forward region.  More than $1.3\times 10^6$ events from clean
low-luminosity runs with the detector functioning properly have been analyzed
for
the work presented here. 

With regard to our DCC search, we have seen no evidence for the presence of
DCC.  Robust observables sensitive to the partition of pions into
charged and neutral species have been found
to
good accuracy to not depend upon associated multiplicity or upon the presence
or
absence of a leading nucleon, or of a diffractive proton. These results have
been
obtained from a data-driven analysis method developed by the
collaboration, a
method which has general applicability to DCC searches \cite{WA98}. 

In order to determine the limits on DCC production implied by our measurements,
we have to face two issues:  
the determination of a lower bound on the
possible values of the robust observables consistent with our data; and
the dependence of the robust observables on
various models for DCC production.  

The question of a lower bound on the robust observables is complicated by the
discrepancy between the values resulting from our two tracking algorithms. For
the statistically most significant observable, $r_{1,1}$, the combinatorial
tracker yielded a value of $r_{1,1}=1.026 \pm 0.004$ while the $u-v$ tracker yielded
a value of $r_{1,1}=1.140 \pm 0.009$.  This discrepancy is also present in the 
respective results of the Monte-Carlo simulations:  the combinatorial tracker
yielded a value of $r_{1,1}=1.02\pm 0.02$, while the $u-v$ tracker yielded
$r_{1,1}=1.13 \pm 0.03$.  

We believe that a realistic approach to this discrepancy
is to normalize the results of each tracker on the data to the results on the
Monte-Carlo.  For the combinatorial tracker, this yields 
\begin{equation}
r^{normalized}= {r^{data}_{1,1}\over r^{MC}_{1,1}} = 1.01 \pm 0.02 
%
\qquad\hbox{(combinatorial tracker)}.
\end{equation}
The corresponding one-sided lower limit at the 95\% confidence level ($1.645 \sigma$)
is $r^{normalized} \geq 0.973$.
The $u-v$ tracker yields
\begin{equation}
r^{normalized}= {r^{data}_{1,1}\over r^{MC}_{1,1}} = 1.01 \pm 0.03
%
\qquad\hbox{({\it u-v} tracker)},
\end{equation}
with a one-sided lower limit at the 95\% confidence level of
$r^{normalized} \geq 0.963$.

\begin{figure}
\epsffile{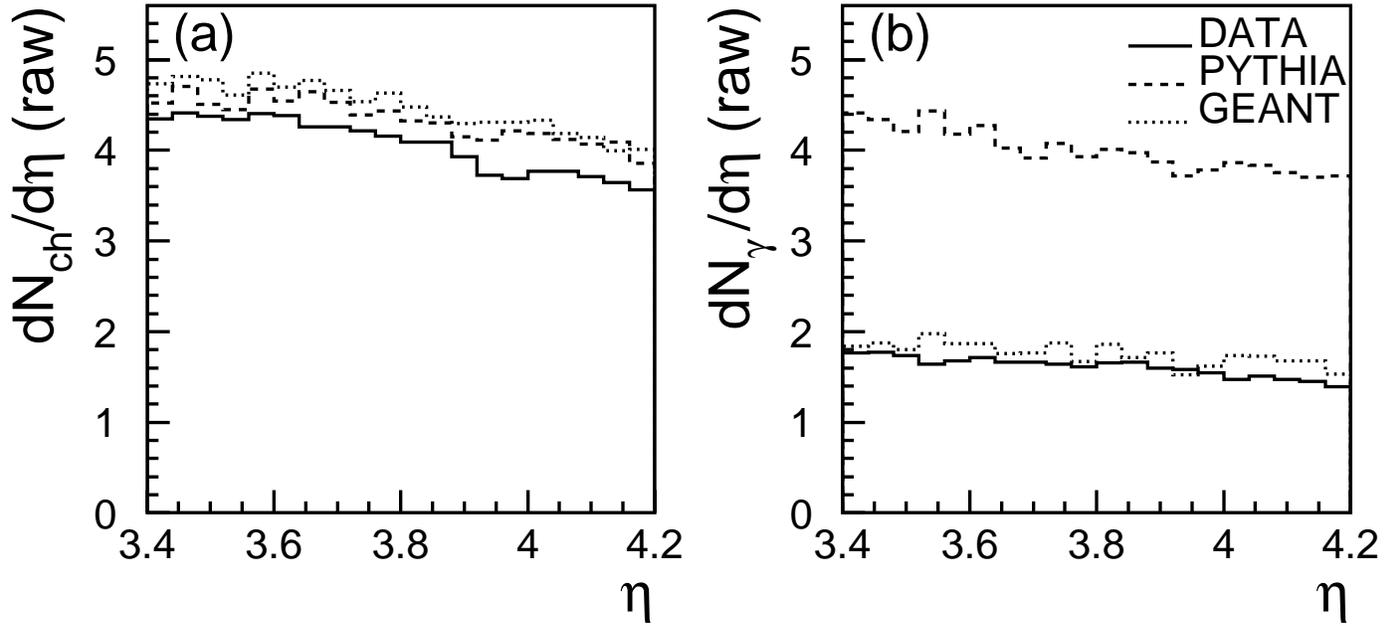}
\caption{FIG. 8 Raw
distributions of the pseudorapidity
distributions of charged and neutral particles. The curve labelled
"PYTHIA" refers to simulated events produced by the PYTHIA event
generator. These events were then propagated through the GEANT
detector simulation and reconstruction algorithms.  These results
are labelled "GEANT".  The large shift between "PYTHIA" and "GEANT"
for the photon distribution is largely due to the conversion probability.
The observed data, uncorrected for
detection and trigger efficiencies are plotted as "DATA".
The close agreement between the "DATA" and
"GEANT" curves validates the simulation procedure and our understanding
of the analysis and cuts.}
\label{dndeta}
\end{figure}

These limits are consistent with generic, binomial-distribution partition of pions
into charged and neutral species.  While the robust observables, being independent
of detection efficiencies, do not permit the determination of the neutral fraction, 
we note that Figure \ref{dndeta} indicates that the normalization of 
the observed inclusive measurements of
gammas and charged particles agree with Pythia/GEANT simulations at roughly the 
10\% level.

The limits on DCC production implied by these lower bounds on the robust observables
are strongly dependent on the model for DCC production one uses.  The nature of
this model dependence has been described in detail in our previous paper 
\cite{robust}.  Two extreme cases illustrate some of the issues: one (``exclusive
production") where a given event is described by either DCC or generic production,
and the other (``associated production") where DCC production is proportional
to generic production.  

In the case of associated production,
\begin{equation}
r^{assoc}_{1,1}(\lambda)=
{[(1-\lambda)^{2}{\hat f}(1-{\hat f}) +
{1\over 3}\lambda (1-\lambda) (1 +{\hat f}) +{2\over 15} \lambda^2]
[(1-\lambda)(1-{\hat f}) +{2\over 3}\lambda ]\over [(1-\lambda)^{2}(1-{\hat
f})^2 +
{4\over 3}\lambda (1-\lambda) (1 -{\hat f}) +{8\over 15}
\lambda^2][(1-\lambda){\hat f} +{1\over 3}\lambda ]},
\end{equation}
where $\lambda = <N>_{DCC}/<N>_{generic}$ and $\hat f$ is the mean fraction of
$\pi^0$'s in generic production, which we take here to be $1/3$.  In this case,
one can solve for an upper limit on $\lambda$ in terms of 
$r_{1,1}^{lower\, limit}$:
\begin{equation}
\lambda \leq \sqrt{ 5 (1-r_{1,1}^{lower\, limit})\over r_{1,1}^{lower\, limit}+2}.
\end{equation}
For this model of DCC production, we thus find 
an upper limit on 
DCC production of $\lambda \leq 0.21$ 
based on the results of the
the combinatorial tracker.

In the case of exclusive production, 
\begin{equation}
r^{excl}_{1,1}(\lambda)=
{ [1+ \lambda({2\over 15 {\hat f}(1-{\hat f})} {\langle N(N-1)\rangle^{DCC}
\over \langle N(N-1)\rangle^{Gen}} -1)] [1+ \lambda({2\over 3 (1-{\hat f})}
{\langle N)\rangle^{DCC} \over \langle N\rangle^{Gen}} -1)] \over [1+
\lambda({8\over 15 (1-{\hat f})^2}
{\langle N(N-1)\rangle^{DCC} \over \langle N(N-1)\rangle^{Gen}} -1)][1+
\lambda({1\over 3 {\hat f}}
{\langle N)\rangle^{DCC} \over \langle N\rangle^{Gen}} -1)] }, 
\end{equation}
where $\lambda$ is the probability that an event will be DCC.
In general, this expression
depends on ratios of the first and second moments of the
parent DCC and generic multiplicity distributions, as well as on $\hat f$.
Illustrative bounds follow from assuming that the parent distributions are
the same, and that $\hat f = 1/3$.  In this case, one can again solve for an
upper limit on $\lambda$ in terms of our lower limits on $r_{1,1}$:
\begin{equation}
\lambda \leq { 5 (1-r_{1,1}^{lower\, limit})\over r_{1,1}^{lower\, limit}+2}.
\end{equation}
For this model of DCC production, we thus find 
an upper limit on
DCC production of $\lambda \leq 0.05$.

A third model, independent production, in which DCC production occurs 
independently of generic production, was also discussed in our earlier
paper.  As noted there, the precise value of the robust observables depends
in this case on the details of both generic and DCC production, so that
analytic formulae such as those we have just considered are of less
utility.  In our earlier paper, we have reported the results for a series
of Monte-Carlo simulations in which a varying amounts of DCC are added to
generic events.  The parameters of the DCC generator used for these simulations
corresponded to domains of DCC with energy density comparable to that
of generic production, and $<p_t>\sim \hbox{140 MeV}$. Interpolating from
the results reported in Table II of \cite{robust}, we deduce 
limits on DCC production in this scenario of about 5\%.

Similar analyses (and conclusions)
are possible for the data subsets defined by the diffractive
and forward antinucleon tags, and for events with opposite side multiplicity tags.
Indeed, there is no evidence of a multiplicity dependence in the robust observables.
(See Figure \ref{rpbar}).  
We believe that this, together with the overall agreement between
data and simulations suggested by Figure \ref{dndeta}, 
goes far towards validating our results.

Detailed discussion of the third and fourth goals of the experiment is beyond
the
scope of this paper. We have of course made preliminary searches for exotic
phenomena, and have seen no evidence for them.  A major reason for not
reporting
them  here in more detail is that about 1\% of our triggers have so many
hits in
the chambers (e.g. $N_{hits} > 1200$, or a mean occupancy per wire of more than
40\%)
that the event is almost completely unanalyzable. One may therefore argue that
the
exotic events might be only present in this subset of data. We have some
circumstantial evidence against this argument, because in many of these very
cluttered events there are portions of the acceptance which are relatively
clean,
and in those regions no unusual behavior is indicated. 

Were Centauro behavior present, we would expect to see an excess of events with
high charged multiplicity; no tail in the multiplicity distribution is
observed.
The JACEE collaboration has exhibited a highly unusual event containing high
multiplicity and a gamma ray excess \cite{JACEE}. Our joint distribution in gammas and
charged
tracks significantly limits this behavior, provided the gamma ray efficiency as
determined in Figure~\ref{effnhits} can be extrapolated further in
multiplicity. But we again emphasize that in all these cases these limits would
be
only valid if the background problem described in the previous paragraph is
ignored.

With regard to the fourth goal of the experiment, measurements of
$dN_{ch}/d\eta$
and $dN_\gamma/d\eta$ have been made in a previously unexplored region of phase
space (c.f. Figure \ref{dndeta}). 
Fully normalized distributions, which take into account detection and
trigger efficiencies and fake tracks, have not yet been derived. Further work
on
modeling fake tracks in the data is necessary before this can be done.

This test/experiment was an extremely modest endeavor. Many lessons were
learned,
which may apply to more ambitious efforts in the future. Some are obvious: it
would be beneficial to have larger acceptance, together with momentum
measurement
of both charged particles and gammas. With regard to the serious backgrounds in
the forward direction, miniaturization of the detection elements with
finer spatial resolution, in both
tracking
and in calorimetry, greatly reduces the difficulties. An example of a 
proposed detector
with such properties is the FELIX initiative for the LHC\cite{FELIX}.

\section{Acknowledgements}

The MiniMax collaboration  gratefully acknowledges the 
superb support we uniformly received from the Fermilab staff.  We are
similarly grateful for support we received from the Fermilab experimental
community.  We would also like to thank  W. J. Fickinger, L. H. Hinkley, 
R. A. Leskovec and Steven Jogan for specific contributions at CWRU, and
Brenda Kirk for her contributions throughout the project. 

This work was supported in part by the U.S. Department of Energy, the U.S.
National
Science Foundation, the Guggenheim Foundation, the Timken Foundation, 
the Ohio Supercomputer Center, and the College of Arts and Sciences and
the Provost's Fund of
Case Western Reserve University.


%
%

\vfill\eject

%
%

\begin{table}[h]
\caption{The number of events observed with a given 
$n_{ch}$, $n_\gamma$: minimum bias data set.}
\label{t:ncg_comp}
\begin{tabular}{|cc|cccccccccl|}
\hline
 & & \multicolumn{9}{c}{$n_\gamma$}  & \\
 & & \ \ 0\ \ \  & \ \ 1\ \ \  & \ \ 2\ \ \  & \ \ 3\ \ \  & \ \ 4\ \ \  &
\ \ 5\ \ \  &\ \ 6\ \ \  & \ \ 7\ \ \ &\ \ 8\ \ \ & \\
\hline
 & 0 & 742039 & 111976 & 14868  & 2237  &  401  &   60 &  9 & 8 & 1 & \\
 & 1 & 318521 &  56182 &  9324  & 1557  &  258  &   49 & 11 & 1 & 0 & \\
 & 2 &  78220 &  17232 &  3204  &  573  &  139  &   27 &  3 & 0 & 0 & \\
 & 3 &  16018 &   4321 &   912  &  184  &   32  &    8 &  0 & 0 & 0 & \\
$n_{ch}$ & 4 & 3021 & 856 & 197  &  43  &    8  &    2 &  0 & 0 & 0 & \\
 & 5 &    473 &    170 &    46  &    7  &    4  &    0 &  0 & 0 & 0 & \\
 & 6 &     76 &     34 &     5  &    2  &    0  &    0 &  0 & 0 & 0 & \\
 & 7 &     10 &      5 &     1  &    0  &    0  &    0 &  0 & 0 & 0 & \\
 & 8 &      0 &      1 &     0  &    0  &    0  &    0 &  0 & 0 & 0 & \\
\hline
\end{tabular}
\end{table}

\begin{table}[h]
\caption{The observed number of events with a given 
$n_{ch}$, $n_\gamma$ in which there is a leading antiproton
of $x\sim 0.9$.}
\label{t:ncg_ktag}
\begin{tabular}{|cc|cccccccccl|}
\hline
 & & \multicolumn{9}{c}{$n_\gamma$}  & \\
 & & \ \ 0\ \ \  & \ \ 1\ \ \  & \ \ 2\ \ \  & \ \ 3\ \ \  & \ \ 4\ \ \  &
\ \ 5\ \ \  &\ \ 6\ \ \  & \ \ 7\ \ \ &\ \ 8\ \ \ & \\
\hline
 & 0 & 11224 &  1532 &   188 &    25 &     7 &     0 & 0 & 0 & 0 & \\
 & 1 &  4245 &   622 &    97 &    10 &     2 &     1 & 0 & 0 & 0 & \\
 & 2 &   798 &   173 &    27 &     8 &     0 &     0 & 0 & 0 & 0 & \\
 & 3 &   145 &    39 &     5 &     2 &     0 &     0 & 0 & 0 & 0 & \\
$n_{ch}$ & 4 & 13 & 5 &    2 &     0 &     1 &     0 & 0 & 0 & 0 & \\
 & 5 &     6 &     2 &     1 &     0 &     0 &     0 & 0 & 0 & 0 & \\
 & 6 &     0 &     0 &     0 &     0 &     0 &     0 & 0 & 0 & 0 & \\
 & 7 &     0 &     0 &     0 &     0 &     0 &     0 & 0 & 0 & 0 & \\
 & 8 &     0 &     0 &     0 &     0 &     0 &     0 & 0 & 0 & 0 & \\
\hline
\end{tabular}
\end{table}

\begin{table}[h]
\caption{The observed number of events with a given 
$n_{ch}$, $n_\gamma$ in which there is a leading antiproton
of $x\sim 0.5$.}
\label{t:ncg_pbar}
\begin{tabular}{|cc|cccccccccl|}
\hline
 & & \multicolumn{9}{c}{$n_\gamma$}  & \\
 & & \ \ 0\ \ \  & \ \ 1\ \ \  & \ \ 2\ \ \  & \ \ 3\ \ \  & \ \ 4\ \ \  &
\ \ 5\ \ \  &\ \ 6\ \ \  & \ \ 7\ \ \ &\ \ 8\ \ \ & \\
\hline
 & 0 & 21447 &  3068 &   403 &    58 &    16 &     1 & 0 & 0 & 0 & \\
 & 1 &  8594 &  1439 &   219 &    34 &     5 &     0 & 0 & 0 & 0 & \\
 & 2 &  1795 &   391 &    83 &    15 &     2 &     0 & 0 & 0 & 0 & \\
 & 3 &   334 &    80 &    23 &     2 &     2 &     0 & 0 & 0 & 0 & \\
$n_{ch}$ & 4 & 61 & 18 &   5 &     1 &     0 &     1 & 0 & 0 & 0 & \\
 & 5 &     5 &     7 &     0 &     0 &     0 &     0 & 0 & 0 & 0 & \\
 & 6 &     2 &     1 &     0 &     0 &     0 &     0 & 0 & 0 & 0 & \\
 & 7 &     0 &     0 &     0 &     0 &     0 &     0 & 0 & 0 & 0 & \\
 & 8 &     0 &     0 &     0 &     0 &     0 &     0 & 0 & 0 & 0 & \\
 \hline
\end{tabular}
\end{table}

\begin{table}[h]
\caption{The observed number of events with a given 
$n_{ch}$, $n_\gamma$ in which there is a leading zero degree
neutral particle.}
\label{t:ncg_nbar}
\begin{tabular}{|cc|cccccccccl|}
\hline
 & & \multicolumn{9}{c}{$n_\gamma$}  & \\
 & & \ \ 0\ \ \  & \ \ 1\ \ \  & \ \ 2\ \ \  & \ \ 3\ \ \  & \ \ 4\ \ \  &
\ \ 5\ \ \  &\ \ 6\ \ \  & \ \ 7\ \ \ &\ \ 8\ \ \ & \\
\hline
 & 0 & 14313 &  1944 &   244 &    37 &     5 &     1 & 0 & 0 & 0 & \\
 & 1 &  5439 &   871 &   139 &    22 &     2 &     2 & 0 & 0 & 0 & \\
 & 2 &  1138 &   232 &    42 &     6 &     0 &     0 & 0 & 0 & 0 & \\
 & 3 &   215 &    41 &     9 &     3 &     0 &     0 & 0 & 0 & 0 & \\
$n_{ch}$ & 4 & 31 & 9 &    1 &     1 &     0 &     0 & 0 & 0 & 0 & \\
 & 5 &     5 &     3 &     2 &     0 &     0 &     0 & 0 & 0 & 0 & \\
 & 6 &     0 &     0 &     0 &     0 &     0 &     0 & 0 & 0 & 0 & \\
 & 7 &     0 &     0 &     0 &     0 &     0 &     0 & 0 & 0 & 0 & \\
 & 8 &     0 &     0 &     0 &     0 &     0 &     0 & 0 & 0 & 0 & \\
 \hline
\end{tabular}
\end{table}

\begin{table}[h]
\caption{Number of events with a given $n_{ch}$, $n_\gamma$ 
from the 10\% of events with lowest energy in the hodoscope  
($E<2.5$ mips).}
\label{t:ncg_pb1}
\begin{tabular}{|cc|cccccccccl|}
\hline
 & & \multicolumn{9}{c}{$n_\gamma$}  & \\
 & & \ \ 0\ \ \  & \ \ 1\ \ \  & \ \ 2\ \ \  & \ \ 3\ \ \  & \ \ 4\ \ \  &
\ \ 5\ \ \  &\ \ 6\ \ \  & \ \ 7\ \ \ &\ \ 8\ \ \ & \\
\hline
 & 0 & 84513 &  9609 &  1034 &   138 &    27 &     4 & 1 & 0 & 0 & \\
 & 1 & 27050 &  3527 &   478 &    68 &    11 &     1 & 0 & 0 & 0 & \\
 & 2 &  4737 &   796 &   126 &    14 &     4 &     0 & 1 & 0 & 0 & \\
 & 3 &   737 &   171 &    22 &     8 &     1 &     0 & 0 & 0 & 0 & \\
$n_{ch}$ & 4 &  98 & 22 &  1 &     1 &     0 &     0 & 0 & 0 & 0 & \\
 & 5 &    14 &     6 &     0 &     0 &     0 &     0 & 0 & 0 & 0 & \\
 & 6 &     4 &     0 &     0 &     0 &     0 &     0 & 0 & 0 & 0 & \\
 & 7 &     0 &     0 &     0 &     0 &     0 &     0 & 0 & 0 & 0 & \\
 & 8 &     0 &     0 &     0 &     0 &     0 &     0 & 0 & 0 & 0 & \\
\hline
\end{tabular}
\end{table}

\begin{table}[h]
\caption{Number of events with a given $n_{ch}$, $n_\gamma$
from the 10\% of events with highest energy in the hodoscope 
($E>34$ mips).}
\label{t:ncg_pb10}
\begin{tabular}{|cc|cccccccccl|}
\hline
 & & \multicolumn{9}{c}{$n_\gamma$}  & \\
 & & \ \ 0\ \ \  & \ \ 1\ \ \  & \ \ 2\ \ \  & \ \ 3\ \ \  & \ \ 4\ \ \  &
\ \ 5\ \ \  &\ \ 6\ \ \  & \ \ 7\ \ \ &\ \ 8\ \ \ & \\
\hline
 & 0 & 59926 & 11399 &  1748 &   281 &    55 &    10 & 0 & 2 & 0 & \\
 & 1 & 33686 &  7668 &  1450 &   275 &    36 &     7 & 1 & 0 & 0 & \\
 & 2 & 10924 &  2905 &   595 &   117 &    24 &     7 & 0 & 0 & 0 & \\
 & 3 &  2705 &   851 &   169 &    39 &    10 &     1 & 0 & 0 & 0 & \\
$n_{ch}$ & 4 & 575 & 183 & 42 &    7 &     0 &     0 & 0 & 0 & 0 & \\
 & 5 &   112 &    31 &    10 &     1 &     0 &     0 & 0 & 0 & 0 & \\
 & 6 &    16 &    11 &     1 &     0 &     0 &     0 & 0 & 0 & 0 & \\
 & 7 &     2 &     1 &     0 &     0 &     0 &     0 & 0 & 0 & 0 & \\
 & 8 &     0 &     0 &     0 &     0 &     0 &     0 & 0 & 0 & 0 & \\
\hline
\end{tabular}
\end{table}

\begin{table}[h]
\caption{Values of $r_{i,j}$ for all Pb-in events and for
those with upstream tags.  
Only the $r_{i,1}$ are robust; the other quantities are tabulated for completeness.
The entries in a given column are not statistically independent.
The last three rows are calculated using the
$u-v$ tracker; all others are results from the combinatorial tracker.
$\# $ events refers to the number of raw events put through the respective
trackers.}
\label{t:rreal}
\begin{tabular}{|c|c|c|c|c|}
\hline
 & all events & diffractive-$\bar p$ & forward-$\bar n $ & 
 forward-$\bar p$\\
\hline
 \# events & 1383336 & 19180 & 24757 & 38112 \\
\hline
$\left< n_{ch}\right>$ & $\quad 0.4843\pm 0.0006\quad$ & 
  $\quad 0.401\pm 0.005\quad$ & $\quad 0.417\pm 0.004\quad$ & 
  $\quad 0.436\pm 0.004\quad$ \\
$\left< n_{\gamma}\right>$ & $0.1923\pm 0.0004$ & 
   $0.166\pm 0.003$ & $0.171\pm 0.003$ & $0.181\pm 0.002$ \\
$\left< n_{ch}(n_{ch}-1)\right>$ & $0.2858\pm 0.0010$ &
   $0.187\pm 0.007$ & $0.208\pm 0.006$ & $0.225\pm 0.005$ \\
$\left< n_{ch}n_{\gamma}\right>$ & $0.1165\pm 0.0005$ &
   $0.083\pm 0.003$ & $0.089\pm 0.003$ & $0.100\pm 0.003$ \\
\hline
 $r(1,1)$ & $1.026 \pm 0.004$ & $1.07 \pm 0.04$
          & $1.04 \pm 0.03$ & $1.07 \pm 0.03$ \\
 $r(2,1)$ & $1.035 \pm 0.010$ & $1.17 \pm 0.12$
          & $1.05 \pm 0.09$ & $1.11 \pm 0.07$ \\
 $r(3,1)$ & $1.059 \pm 0.027$ & $1.34 \pm 0.32$
          & $1.29 \pm 0.22$ & $1.18 \pm 0.18$ \\
 $r(4,1)$ & $1.118 \pm 0.065$ & $1.77 \pm 0.66$
          & $2.40 \pm 0.68$ & $1.38 \pm 0.44$ \\
 $r(5,1)$ & $1.310 \pm 0.151$ &
          &                   & $1.74 \pm 0.92$ \\
 $r(6,1)$ & $1.904 \pm 0.382$ &
          &                   &                   \\
\hline
 $r(0,2)$ & $1.586 \pm 0.010$ & $1.70 \pm 0.10$
          & $1.66 \pm 0.08$ & $1.66 \pm 0.06$ \\
 $r(1,2)$ & $1.573 \pm 0.022$ & $1.70 \pm 0.25$
          & $1.63 \pm 0.17$ & $1.65 \pm 0.16$ \\
 $r(2,2)$ & $1.556 \pm 0.053$ & $2.08 \pm 0.73$
          & $1.71 \pm 0.40$ & $1.87 \pm 0.47$ \\
 $r(0,3)$ & $3.420 \pm 0.082$ & $3.64 \pm 0.72$
          & $3.47 \pm 0.63$ & $3.31 \pm 0.41$ \\
 $r(1,3)$ & $3.165 \pm 0.129$ & $3.49 \pm 1.35$
          & $2.98 \pm 0.88$ & $3.16 \pm 0.99$ \\
 $r(0,4)$ & $9.251 \pm 0.683$ & $7.65 \pm 3.45$
          & $8.56 \pm 3.82$ & $6.11 \pm 1.79$ \\
\hline
 \# events & 242959 & 22477 & 29026 & 44404 \\
\hline
 $r(1,1)_{uv}$ & $1.140 \pm 0.009$ & $1.17 \pm 0.04$
          & $1.18 \pm 0.03$ & $1.19 \pm 0.02$  \\
 $r(2,1)_{uv}$ & $1.281 \pm 0.026$ & $1.33 \pm 0.11$
          & $1.45 \pm 0.09$ & $1.35 \pm 0.08$  \\
 $r(3,1)_{uv}$ & $1.506 \pm 0.076$ & $1.62 \pm 0.32$
          & $1.41 \pm 0.35$ & $1.36 \pm 0.22$  \\
\hline
\end{tabular}
\end{table}

\begin{table}[h]\vspace{0.75in}
\caption[Mean multiplicities and robust observables for bins
of hodoscope multiplicities each containing 10\% of the events.]
{Mean multiplicities and robust observables for bins
of hodoscope multiplicities each containing 10\% of the events.  
The hodoscope multiplicity increases with increasing bin number 1-10.}
\label{t:rpb}
\begin{center}
\begin{tabular}{|c|c|c|c|c|c|}
\hline
 bin & $\left< n_{ch}\right>$ & $\left< n_\gamma\right>$ &  
 $r_{1,1}$   & $r_{2,1}$    & $r_{3,1}$ \\
\hline
1 & $\quad 0.345\pm 0.002\quad$ & $\quad 0.138\pm 0.001\quad$ & 
  $\quad 1.03\pm 0.02\quad$ & $\quad 1.01\pm 0.05\quad$ & 
  $\quad 0.91\pm 0.13\quad$ \\
2 & $0.390\pm 0.002$ & $0.159\pm 0.001$ & 
  $1.03\pm 0.01$ & $1.07\pm 0.04$ & $1.22\pm 0.13$ \\
3 & $0.419\pm 0.002$ & $0.168\pm 0.001$ & 
  $1.04\pm 0.01$ & $1.07\pm 0.04$ & $1.13\pm 0.10$ \\
4 & $0.447\pm 0.002$ & $0.179\pm 0.001$ & 
  $1.03\pm 0.01$ & $1.04\pm 0.04$ & $1.06\pm 0.10$ \\
5 & $0.471\pm 0.002$ & $0.189\pm 0.001$ & 
  $1.03\pm 0.01$ & $1.05\pm 0.04$ & $1.10\pm 0.10$ \\
6 & $0.499\pm 0.002$ & $0.196\pm 0.001$ & 
  $1.03\pm 0.01$ & $1.07\pm 0.04$ & $1.17\pm 0.09$ \\
7 & $0.516\pm 0.002$ & $0.206\pm 0.001$ & 
  $1.03\pm 0.01$ & $1.06\pm 0.03$ & $1.11\pm 0.09$ \\
8 & $0.549\pm 0.002$ & $0.218\pm 0.001$ & 
  $1.01\pm 0.01$ & $1.01\pm 0.03$ & $1.08\pm 0.07$ \\
9 & $0.587\pm 0.002$ & $0.230\pm 0.001$ & 
  $1.02\pm 0.01$ & $1.00\pm 0.03$ & $0.96\pm 0.06$ \\
10 & $0.646\pm 0.002$ & $0.249\pm 0.001$ & 
  $1.04\pm 0.01$ & $1.04\pm 0.02$ & $1.03\pm 0.05$ \\
\hline
\end{tabular}
\end{center}

\end{table}
\clearpage

\end{document}